# TWIN SORT TECHNIQUE

Veeresh D[1], Dr.Thimmaraju S. N[2], Ravish G. K[3]

[1]Cognizant Tech Solutions, Apple INC San Jose, California, USA.
[2]Professor, Dept. of MCA, PG Center Mysore, India
[3]Assitant.Prof, Dept. of MCA, PG Center Mysore, India

**Abstract** : The objective behind the "TWIN SORT" technique is to sort the list of unordered data elements efficiently and to allow efficient and simple arrangement of data elements within the data structure with optimization of comparisons and iterations in the sorting method. This sorting technique is effectively terminates the iterations when there is no need of comparison if the elements are all sorted in between the iterations. Unlike Quick sort, Merge sorting technique, this new sorting technique is based on the iterative method of sorting elements within the data structure. So it will be advantageous for optimization of iterations when there is no need for sorting elements. Finally, the "TWIN SORT" technique is more efficient and simple method of arranging elements within a data structure and it is easy to implement when comparing to the other sorting technique. By the introduction of optimization of comparison and iterations, it will never allow the arranging task on the ordered elements.

## I. INTRODUCTION

The Sorting is an important concept of programming. The sorting method allows different ways to arrange the given elements in a data structure. There are different sorting methods with their own best-case and worst-case efficiencies. But in almost all sorting techniques there are no optimizations for preventing and reducing the unnecessary comparisons or iterations. So in order to allow an effective and simple arrangement of elements with least number of comparisons and iterations within a data structure, the new iterative sorting technique "Twin Sort" is introduced. This sorting technique possesses a good average case behavior with effective optimizations in comparisons and iterations during arranging the data elements. The basic terminology behind this sorting technique is the Twin or couple of elements. This algorithm works by considering (dividing) array elements as Twins of elements and the elements which comes in these Twins (every two elements) are sorted iteratively with existence of optimization in between the iterations i.e., here the optimization will terminate the sorting process if all the elements within these Twins are sorted within N iterations. Therefore here the way we are optimizing the iterations and its comparisons will lead to the excellent time and memory space efficiency. In the "Twin Sort" method of sorting elements, over all tasks is divided according to the EVEN and ODD number of iterations. Because, for the purpose of swapping the Twin‟s elements, the formation of Twins (consideration of multiple two elements) must be done at two particular starting positions they are at [0] th (first) position and [0+1] th (second) position in the array depending on the EVEN and ODD number of the iteration. Here we usually start the iterations with number from 0 to N-1 (i.e., we start the loop from 0 to N-1). If the iteration number is EVEN (say 0,2,4,6, etc) then we start forming a Twin of elements (considering two elements) sequentially starting at 0th position. Therefore the first Twin will be (array[0],array[1]), the second Twin will be (array[2],array[3]) and the third will be (array[4], array[5]) and so on. Similarly, if the iteration number is ODD (say 1,3,5,7, etc) then we start forming a Twin of elements (considering two elements) sequentially starting at [0+1] th i.e., at 2nd position, therefore the first Twin will be (array [1], array [2]) the second Twin will be (array [3], array [4]) and the third will be (array [5], array [6]) and so on. As and when we form the Twin elements, we swap the elements of the Twin, if there is difference between those two elements of the Twin, else we keep those Twin elements as it is in the array or data structure. In this case no swapping operation is done. During the determination of the TRUE or FALSE conditions of the differences of Twin elements, we then keep count of FALSE conditions, this count value will be used to terminate the sorting process and exit from it to optimize the further comparisons. Importantly, the sorting process is terminated when the „Count of FALSE conditions is equal to the (N-1)" where N is the number of input elements. In this new sorting technique, all iteration needs exactly „N/2" comparisons, if there is ODD number of input elements i.e., if N is ODD number. Similarly, it needs 'N/2' comparisons in ith iteration and (N-1)/2 comparisons in (i-1)th iteration if there are EVEN number of elements i.e., if N is even.

## II. Algorithms

1. Read input size „n‟ and set of data elements
2. Make a loop while not equal to number of elements „n‟ 2.1 Make a loop while not equal to „n / 2‟
    a) Form a new pair elements
    b) IF difference exist in pair elements THEN (i.e., True condition)
       Swap pair elements.
   ELSE (Count the number of FALSE conditions)

c) IF pair elements are all in order (when is n equal to no.of False conditions-1) THEN Terminate sorting process and exit function.
3. End.

### III. Analysis with Example

**Analysis with Example**

Let us illustrate the "Twin Sort" technique using the following Descendant input elements to be sorted into ascendant elements:
**Input elements: 5 , 4 , 3 , 2 , 1**

**Initial Data structure (Array structure):**

Input elements Swapped elements Array
elements Iteration 0: (5, 4) (3, 2), 1 (4, 5) (2, 3), 1
**4 5 2 3 1**

Iteration 1: 4, (5, 2) (3, 1) 4, (2, 5) (1, 3)
**4 2 5 1 3**

Iteration 2: (4, 2) (5, 1), 3 (2, 4) (1, 5), 3
**2 4 1 5 3**

Iteration 3: 2, (4, 1) (5, 3) 2, (1, 4) (3, 5)
**2 1 4 3 5**

Iteration 4: (2, 1) (4 3), 5 (1, 2) (3, 4), 5
**1 2 3 4 5**

Analysis of Efficiencies
In the analysis of efficiencies of this sorting algorithm, two kinds of efficiencies must be considered. They are:
1. Time efficiency
2. Space efficiency

Time efficiency measures how fast an algorithm runs and the amount of time it needs. This new method of sorting consumes less time by eliminating n/2 comparisons i.e., it makes only n/2 comparisons in a single iteration. As a best-case efficiency it takes a maximum of (n-1) comparisons in two iterations. This best-case efficiency is the outstanding property of this sorting technique, because of the constant factors (n-1) and two iterations. Space efficiency is related to the amount of memory required to sort the elements and the amount of extra memory the algorithm needs. In this new sorting technique, it does not use any extra memory space in sorting elements. This sorting technique being iterative method does not use the stack memory unlike recursive based sorting methods (Quick, Merge etc). And primarily, this sorting technique is mainly based on individual Twin elements (two elements only); the swapping of elements will be done within the data structure. According to the standard notations of algorithms efficiencies, I have derived the following efficiencies for this new sorting algorithm:

Best-case efficiency:
For all n: **C (n) = O (n-1)** f(n)= n-1 eg: f(25)=5-1 =4

Worst-case efficiency:
For all n: **C (n) = O ((n-1)*(n/2))** ⌐ **O (n Log n)** f(n)=(n-1)*n/2 eg: f(5)=(5-1)*5/2 =10

Average case efficiency:
For all n: **C (n) = O ([(n-1)*(n/2)]/2)** ⌐ **O (n Log n)** f(n)=[(n-1)*n/2]/2 eg: f(5)=[(5-1)*5/2]/2 =5

The above content of table represents the Best case, Worst-case and Average case efficiencies of this new sorting technique. Those efficiencies are practically obtained values by giving the input elements ranging up to ten elements (unordered). And by analyzing the outputs of this sorting technique with those input elements, I have derived the efficiencies in terms of Best, Worst and Average cases. Importantly, I have derived the efficiencies (values), which indicate the basic operations, which are involved in the arrangement of Twin"s elements that include: number of comparisons, which are used for it. In the Best-case efficiency, the function: f (n) =n-1 gives the best case efficiency of this algorithm, its each values includes the (n-1) number of basic operations, where the value (n-1) indicates the maximum number of comparisons to be used in the iterations (loops). Similarly, in the Worst-case efficiency, the function: f (n) = (n-1)*n/2 gives the worst-case efficiency of this algorithm. Its each values includes the (n-1)*n/2 number of basic operations, where the value „n/2" indicates the number of comparisons to be done in every iteration (loop).

**Scope and Life**
There are plenty of applications and uses of the sorting technique, they are exist in the field of computer science and engineering technology. Some of the important applications of the sorting technique are as mentioned below:
- For Sorting data elements or names of
- entities In Binary Searching Technique

- For tuple or records ordering in the tables of DBMS.

- In the field of operating systems, for ordering the files and directory structure

- In the field of Graph Theory.In the field of communication network, as a

Frame sorting method at the receiver side usually

### IV Practical Efficiencies

| Input | Best Case Efficiency | Worst Case Efficiency | Avg. Case Efficiency |
|---|---|---|---|
| n | f(n)=n-1 | f(n)=(n-1)*n/2 | f(n)=[(n-1)*n/2]/2 |
| 2 | 1 | 1 | 0.5 |
| 3 | 2 | 3 | 1.5 |
| 4 | 3 | 6 | 3 |
| 5 | 4 | 10 | 5 |
| 6 | 5 | 15 | 7.5 |
| 7 | 6 | 21 | 10.5 |
| 8 | 7 | 28 | 14 |
| 9 | 8 | 36 | 18 |
| 10 | 9 | 45 | 22.5 |
| 11 | 10 | 55 | 27.5 |
| 12 | 11 | 66 | 33 |
| 13 | 12 | 78 | 39 |

## V. Conclusion

This new sorting technique being iterative method of arranging the data elements in to the ordered elements, this would consumes less time by eliminating the unnecessary iterations and optimizing the comparisons in arranging the elements in an iterative manner unlike recursive based method of sorting elements. Here the task of arranging elements being within a data structure, it does not consume extra memory space (stack memory) as used by the recursive based sorting techniques (Quick sort, Merge sort etc.,) in keeping track of its calls and its return values. Finally, the "Twin Sort" technique is more efficient and simple method of arranging elements within a data structure and it is easy to implement when comparing to the other sorting technique. By the introduction of optimization of comparison and iterations, it will never allow the arranging task on the ordered elements.

## Acknowledgements

The authors would like to acknowledge the support received from the Visvesvaraya Technological University, Belgaum. Thanks are also to the Department of Mathematics and Department of Computer Science for their co-operation.